\documentstyle{article}
\Roman{section}

\begin{document}
\begin{titlepage}
\centerline{\large \bf INELASTIC RESCATTERING AND CP ASYMMETRIES 
IN $D\to\pi^+\pi^-,\; \pi^0\pi^0$ 
\footnote{This work is partly supported by the National Natural Science
Foundation of China (NSFC) }}

\vspace{2.5cm}

\centerline{Xue-Qian Li$^{1,2}$, and Bing-Song Zou$^{3}$}
\vspace{1cm}
{\small
\begin{center}

1.   CCAST  (World Laboratory)
P.O.Box 8730 Beijing 100080, China.
\vspace{8pt}

2. Department of Physics, Nankai University,
Tianjin, 300071, China.
\vspace{8pt}

3. Queen Mary and Westfield College, London E1 4NS, UK
\end{center}
}
\vspace{2cm}

\begin{center}
{\bf Abstract}
\end{center}
\vspace{1cm}

We study the direct CP violation induced by inelastic final state interaction
(FSI) rescattering in $D\to\pi\pi$ modes, and find that the resultant CP
asymmetry is about $10^{-4}$ which is larger than
$\epsilon'$ in the K-system. Our estimation is based on well-established
theories and experiment measured data, so there are almost no free
parameters except the weak phase $\delta_{13}$ in the CKM matrix.

\vspace{1cm}
PACS number(s): 11.30.Er, 13.25.Ft, 13.25.-k, 13.75.Lb

\end{titlepage}

\noindent{\bf I. Introduction.}
\vspace{0.3cm}

The study on CP violation phenomena is so important for understanding the
mechanisms in particle physics that the subject has attracted great
attention of both experimentalists and theoreticians of high energy physics
for several decades already. The first and so far the only observation of
CP violation is the measurement of $\epsilon$ in the neutral K-system, as
$\epsilon\sim 2.3\times 10^{-3}$ \cite{Data}. the non-zero $\epsilon$ is
due to a mixing of $K^0-\bar K^0$ through box diagrams, which
is theoretically  evaluated at the quark level and the result predicts
an approximate mass of charm-quark \cite{Lee}. However, the parameter
$\epsilon'$ which is related to processes as a direct CP violation has not
been reliably measured yet, it is believed that $\epsilon'/\epsilon$
is non-zero and is expected between $10^{-4}$ to $3\times 10^{-3}$
\cite{Data}.

In the well-known theory,
the mechanism for resulting in non-zero $\epsilon'$ is due to an interference
of two pions in the neutral K-decays \cite{Wu}, and all the details are
given in ref.\cite{Chau}. Since 2$\pi$ can be in both I=0 and 2 isospin
eigenstates, if the two isospin channels have different weak and strong
phases, their interference would result in a CP asymmetry proportional to
$\sin(\delta_0-\delta_2)\sin(\theta_0-\theta_2)$ where $\delta_I$ is the weak
phase and $\theta_I$ corresponds to the the phase shifts emerging in
the strong interaction rescattering processes of
I=0 and 2 channels respectively.

Generally, a direct CP violation must be realized through interferences between
at least two amplitudes having different weak and strong phases, no matter
at quark or hadron levels. The weak phase is determined by an underlying 
theory, for
example, the CKM theory \cite{Kab}, two-Higgs-doublet model \cite{Wein}
etc., while the strong phase can be either produced in the strong scattering
at hadron level as in K-system \cite{Chau}, or occur  at
the quark level. If an absorptive part of loops exists the strong
phase is non-zero. For instance, at the high energy processes, B-physics
\cite{Du}, top-physics \cite{Gr} and high energy collisions \cite{Ber}, the
main part of the strong phases come from the absorptive part of loops,
even though at the hadronization process, the hadron rescattering can also
cause a strong phase.

As the observation concerns hadron products, the strong final state
interaction is not negligible \cite{Don1} and 
CP asymmetry may occur due to the phase
shifts in the rescattering. Recently, by studying single pion exchange
inelastic FSI for $D\to VP$ processes, we pointed out that the 
inelastic strong FSI due to t-channel particle exchange may play important 
roles for producing CP violation in D and B hadronic decays \cite{XQ}. 
A very recent estimation by Blok, Gronau and Rosner
\cite{Blok} shows that the inelastic FSI for $B\to\pi\pi,\; K\bar K$ 
may produce th CP asymmetry as large as $10\sim 20\%$, disregarding
time dependent $B$-$\bar B$ mixing effects.
Due to the GIM mechanism, the $D^0-\bar D^0$ mixing is small compared 
to $B^0-\bar B^0$ \cite{Georgi,Paschos},
even though models beyond three-generation standard model may result in
larger mixing effect \cite{Babu}. Recently, Browder and Pakvasa \cite{Brow}
re-studied experimental implications of large CP violation and final state
interactions (FSI) in a search for $D^0-\bar D^0 $ mixing and they
concluded that FSI is important. Close and Lipkin \cite{Close} studied
possible strong FSI due to exotic resonances in D exclusive decays. 
They constrain their analysis at the
most Cabibbo favored channels where the rescattering is elastic only.
Obviously, it would be interesting to study CP
asymmetry in D-system due to the inelastic strong FSI.

In $D\to \pi\pi$ decays, there are both elastic and
inelastic rescatterings, in general, an amplitudes of
$D\to\pi\pi$ should be
\begin{equation}
\label{dtof}
T(D\to f)=T^{tree}(D\to f)+T^{FSI}(D\to f)
\end{equation}
and
\begin{equation}
T^{FSI}=i\sum_n<f|T|n>\rho_n<n|H_{eff}|D>
\end{equation}
where $|n>$ is a complete set of the strong interaction states satisfying
four-momentum conservation, and $\rho_n$ is the density of state $|n>$.
At $\sqrt s=M_D$ energy region, the most important
intermediate states are $\pi\pi,\; K\bar K,\; \rho\rho,\; K^*\bar K^*,\;
a_1\pi,\; a_2\pi,\; K_1K$.
These intermediate mesons can be on
their mass shell, thus to be real particles. The off-shell contribution
can be attributed into the quark level because the intermediate
particles are virtual. 

In our present paper we only consider
the elastic $\pi\pi\to\pi\pi$ and the inelastic $K\bar K\to\pi\pi$ rescattering 
processes which have experiment measurements \cite{BSZ} at the energy of 
D meson mass and therefore are well constrained. 
This is different from the B meson case, where the inelastic scattering 
$\pi\pi\rightleftharpoons K\bar K$ amplitudes can be only estimated by 
theoretical models \cite{Blok}. 
Contributions from other intermediate states may modify the
results by adding a factor round unity, but are unlikely to change the 
whole scenario and order of magnitude of the CP asymmetry.

Considering the direct tree level transition amplitude, $\pi\pi\to\pi\pi$ and
$K\bar K\to\pi\pi$ FSI, we have
\begin{equation}
T(D\to \pi\pi) \equiv T^{tree}(D\to\pi\pi)+
T^{FSI}(D\to \pi\pi\to\pi\pi)+
T^{FSI}(D\to K\bar K\to\pi\pi).
\end{equation}
Here only the inelastic rescattering from $K\bar K$
intermediate states can induce a direct CP violation, but not the
elastic ones.
Since the tree amplitude of $D^0\to \pi\pi$ and $K\bar K$ have different
weak phases
$Arg(V_{cd}^*V_{ud})$ and $Arg(V_{cs}^*V_{us})$ respectively and the phase
shifts in the inelastic rescattering $K\bar K\rightleftharpoons\pi\pi$ is 
non-zero,
the interference between the two parts $T^{tree}(D\to\pi\pi)$
and $T^{FSI}(D\to K\bar K\to\pi\pi)$ would result in a non-zero
CP violation. The contribution of elastic scattering does not change
the weak phase of $T^{tree}(D\to \pi\pi)$, but they can
cause a strong phase shift to $T^{tree}$. Namely,
$$\tilde T^{tree}(D\to\pi\pi)=T^{tree}(D\to\pi\pi)+T^{FSI}(D\to
\pi\pi\to\pi\pi)=T^{tree}\times f e^{i\theta},$$
where $f$ is a scattering probability amplitude.

Moreover, both $D\to K\bar K$ and $D\to\pi\pi$ are Cabibbo
suppressed modes, so their tree amplitudes have the same order of magnitude.
Even though, one expects that the FSI may change their relative ratios
somehow \cite{Zou}, the order remains the same. Their interference may be
large, since two parts suffer the same Cabibbo suppression.

Our numerical results show that the CP asymmetry can be about
$10^{-4}$.

In next section, we present the formulation for evaluating the CP asymmetries,
in the third section we give the numerical results and the final section is
devoted to our conclusion and discussion. \\

\noindent{\bf II. Formulation}
\vspace{0.3cm}

(i)The tree level amplitudes.

The effective Hamiltonians for $c\to d+u+\bar d$ and $c\to
s+u+\bar s$ are \cite{Abbott}
\begin{equation}
\label{cd}
H_{eff}^{(1)}= {G_F\over\sqrt 2}V_{cd}^*V_{ud}[a_1\bar d\gamma_{\mu}
(1-\gamma_5)c\bar u\gamma^{\mu}(1-\gamma_5)d+a_2\bar d\gamma_{\mu}
(1-\gamma_5)d\bar u\gamma^{\mu}(1-\gamma_5)c]+h.c.,
\end{equation}
and
\begin{equation}
\label{cs}
H_{eff}^{(2)}= {G_F\over\sqrt 2}V_{cs}^*V_{us}[a_1\bar s\gamma_{\mu}
(1-\gamma_5)c\bar u\gamma^{\mu}(1-\gamma_5)s+a_2\bar s\gamma_{\mu}
(1-\gamma_5)s\bar u\gamma^{\mu}(1-\gamma_5)c]+h.c.,
\end{equation}
where the color indices are omitted and $a_1,\; a_2$ are parameters as
\begin{equation}
a_1=c_1+\xi_1 c_2,\;\;\;\;\; a_2=c_2+\xi_2 c_1,
\end{equation}
and
\begin{equation}
\xi_1\equiv{1\over N_c}+{r_1\over 2},\;\;\;\;\; \xi_2\equiv {1\over N_c}
+{r_2\over 2},
\end{equation}
where $r_1$ and $r_2$ correspond to a non-factorizable contribution of the
hadronic matrix elements $<\lambda^a\lambda^a>$ \cite{Cheng}.
\begin{equation}
c_1={c_++c_-\over 2},\;\;{\rm and}\;\; c_2={c_+-c_-\over 2}
\end{equation}
where $c_{\pm}$ can be derived with the renormalization group equation,
numerically at the energy scale of the charm quark,
$$c_1\simeq 1.26,\;\;\;\;\; c_2\simeq -0.51. $$
By fitting data, Cheng obtained $r_1\simeq r_2\sim -0.67$ for $D\to
PP$ decay \cite{Cheng}. Then in our later calculations, we use $a_1=1.26$
and $a_2=-0.51$.

As many authors suggested, we can ignore the contributions from the
W-exchange and annihilation quark diagrams \cite{Wirbel}, so the amplitude
at the tree level can be obtained with the vacuum saturation approximation
and the non-factorization effects are absorbed into the parameters
$r_1$ and $r_2$. We have
\begin{eqnarray}
\label{tree}
<\pi^+\pi^-|H_{eff}^{(1)}|D^0> &=& {G_F\over\sqrt 2}V_{cd}^*V_{ud}a_1f_{\pi}
p_{\pi}^{\mu}<\pi^-|\bar d\gamma_{\mu}(1-\gamma_5)c|D^0> \nonumber \\
&=& {G_F\over\sqrt 2}V_{cd}^*V_{ud}a_1f_{\pi}[f_+^{D\pi}(M_D^2-m_{\pi}^2)
+f_-^{D\pi}m_{\pi}^2],
\end{eqnarray}
where $f_{\pm}^{D\pi}$ are the form factors in D to $\pi$ transition. With
the multi-pole approximation \cite{Xu}
\begin{eqnarray}
\label{multi}
f_+(q^2) &=& F_1(q^2)=F_1(q_m^2)({M_{D^*}^2-q_m^2)\over (M_{D^*}^2-q^2)})^n, \\
f_-(q^2) &=& -({M_D^2-q^2\over M_D^2})F_1(q^2),
\end{eqnarray}
where $M_{D^*}=2.010$ GeV is the nearest pole, $q_m^2\equiv= (M_D-m_{\pi})^2$.
In the following, we take the single-pole approximation as n=1 in
eq.(\ref{multi}).

For $D^0\to \pi^0\pi^0$ and $D^0\to K^+K^-$ we have similar
expressions. Ignoring the W-exchange and annihilation, the tree amplitude
for $D^0\to K^0\bar K^0$ is zero and the transition can only be
realized through the elastic and inelastic FSI rescattering.

The calculated $f_{\pm}^{DK}$ values have been checked by using the method
given in ref.\cite{Roberts} and found they coincide with each other very well.
But Roberts' parameters \cite{Roberts} are obtained by fitting the data of
$D\to K$ transition only, so the results
of $f_{\pm}^{D\pi}$ obtained in the multi-pole
approximation deviate from that calculated in terms of the parameters of
ref.\cite{Roberts} assuming an SU(3) symmetry. Therefore we later take only
the values $f_{\pm}^{D\pi}$ obtained in the multi-pole approximation.

(ii) The elastic and inelastic FSI.

The S-matrix for strong interaction is
\begin{equation}
s_{mn}=\delta_{mn}+2i \sqrt{\rho_m}T_{mn}\sqrt{\rho_n},
\end{equation}
where T-matrix is the non-trivial part determined by the strong
interaction Lagrangian.

It is noted that the $\delta_{mn}$ term corresponds to a no-interaction
scattering transition (or the trivial part of the S-matrix),
so is exactly the ``tree" part of eq.(\ref{dtof}).
For the elastic and the inelastic rescattering contributions, 
the amplitudes read
\begin{eqnarray}
T^{FSI}(D^0\to\pi^+\pi^-\to\pi^+\pi^-)&=&
i<\pi^+\pi^-|T|\pi^+\pi^->\rho_\pi<\pi^+\pi^-|H_{eff}^{(2)}|D^0>,\\
T^{FSI}(D^0\to\pi^0\pi^0\to\pi^+\pi^-)&=&
i<\pi^+\pi^-|T|\pi^0\pi^0>\rho_\pi<\pi^0\pi^0|H_{eff}^{(2)}|D^0>,\\
T^{FSI}(D^0\to K^+K^-\to\pi^+\pi^-)&=&
i<\pi^+\pi^-|T|K^+K^->\rho_K<K^+K^-|H_{eff}^{(2)}|D^0>.
\end{eqnarray}

With help of the isospin analysis, for the elastic scattering,
\begin{eqnarray}
&& \rho_\pi<\pi^+\pi^-|T|\pi^+\pi^->={2\over 3}T_0
e^{i\theta_0}+{1\over 3}T_2e^{i\theta_2},\\
&& \rho_\pi<\pi^+\pi^-|T|\pi^0\pi^0>=
{\sqrt 2\over 3}(-T_0e^{i\theta_0}+T_2e^{i\theta_2}),
\end{eqnarray}
where $T_0,\; \theta_0$ and $T_2,\;\theta_2$ are the
measured scattering amplitudes and phase
shifts of I=0 and I=2 channels \cite{BSZ}, respectively. The transitions 
to the $\pi^0\pi^0$ final state have similar expressions.

The contributions from the elastic FSI of $\pi\pi\to \pi\pi$ can be
absorbed into the tree amplitudes, in our case they do not provide
a different weak phase from the  tree amplitudes,
but result in a strong phase shift.
Including the elastic scattering, the amplitudes for $D^0\to\pi^+
\pi^-$ and $\pi^0\pi^0$ can be written as
\begin{eqnarray}
\label{tilde1}
&&\tilde T^{tree}(D^0\to\pi^+\pi^-)  \nonumber \\
&\equiv& T^{tree}(D^0\to\pi^+\pi^-)+T^{FSI}(D^0\to\pi^+\pi^-
\to\pi^+\pi^-)+T^{FSI}(D^0\to\pi^0\pi^0\to\pi^+
\pi^-) \nonumber\\
&=&{G_F\over\sqrt 2}V_{cd}^*V_{ud}f_{\pi}[f_{+}^{D\pi}(M_D^2-m_{\pi}^2)
+f_{-}^{D\pi}m_{\pi}^2]
\cdot [a_1+({2\over 3}a_1-{\sqrt 2\over 3}a_2)iT_0e^{i\theta_0}
+({1\over 3}a_1+{\sqrt 2\over 3}a_2)iT_2e^{i\theta_2}]\nonumber\\
&& \\
\label{tilde2}
&&\tilde T^{tree}(D^0\to\pi^0\pi^0)  \nonumber \\
&\equiv& T^{tree}(D^0\to\pi^0\pi^0)+T^{FSI}(D^0\to\pi^+\pi^-
\to\pi^0\pi^0)+T^{FSI}(D^0\to\pi^0\pi^0\to\pi^0
\pi^0) \nonumber\\
&=&{G_F\over\sqrt 2}V_{cd}^*V_{ud}f_{\pi}[f_{+}^{D\pi}(M_D^2-m_{\pi}^2)
+f_{-}^{D\pi}m_{\pi}^2] 
\cdot [a_2+({1\over 3}a_2-{\sqrt 2\over 3}a_1)iT_0e^{i\theta_0}
+({2\over 3}a_2+{\sqrt 2\over 3}a_1)iT_2e^{i\theta_2}], \nonumber \\
&&
\end{eqnarray}
where the notation $\tilde T^{tree}$ refers to the tree amplitude modified
by the elastic scattering and
$T_i$, $\theta_i$ are measured values.

For the inelastic scattering, $K\bar K\to\pi\pi$, there are experimental 
measurements \cite{BSZ} 
We can decompose $K^+K^-$ and $\pi^+\pi^-$ in terms of the basis of isospin
as
\begin{eqnarray}
|K^+K^-> &=& {1\over\sqrt 2}[|1,0>+|0,0>]_K;\\
|\pi^+\pi^-> &=& [\sqrt{{2\over 3}}|0,0>+\sqrt{{1\over 3}}|2,0>]_{\pi}.
\end{eqnarray}
Thus
\begin{eqnarray}
\label{ine}
&&T^{inelastic}(D^0\to\pi^+\pi^-) \equiv 
 T^{FSI}(D^0\to K^+K^-\to \pi^+\pi^-)  \nonumber \\
=&& {G_F\over\sqrt 2}V_{cs}^*V_{us}f_Ka_1[f_{+}^{DK}(M_D^2-m_K^2)
+f_{-}^{DK}m_K^2]i\sqrt{{1\over 3}}
T_Ke^{i\theta_K}\sqrt{\rho_K/\rho_\pi}
\end{eqnarray}
where $\rho_K/\rho_\pi=\sqrt{1-4m_K^2/M^2_D}/\sqrt{1-4m^2_\pi/M^2_D}$.

The similar expressions can be easily derived for $T^{FSI}(D^0\to
K^+K^-\to\pi^0\pi^0)$.

(iii) The CP violation.

As long as there are more than two channels with different weak and strong
phases, their interferences can result in a direct CP violation. If
the total transition amplitude $T$ is a superposition of two independent
amplitudes $A_1$ and $A_2$ as
\begin{equation}
T=A_1e^{i\delta_1}e^{i\phi_1}+A_2e^{i\delta_2}e^{i\phi_2},
\end{equation}
while its CP conjugate amplitude is
\begin{equation}
\bar T=A_1e^{-i\delta_1}e^{i\phi_1}+A_2e^{-i\delta_2}e^{i\phi_2}.
\end{equation}
Thus a direct CP asymmetry is defined as
\begin{eqnarray}
\label{dircp}
R &=& {|T|^2-|\bar T|^2\over |T|^2+|\bar T|^2}= \nonumber \\
&=& \frac{2A_1A_2\sin(\delta_1-\delta_2)\sin(\phi_1-\phi_2)}{A_1^2
+A_2^2+2A_1A_2\cos(\delta_1-\delta_2)\cos(\phi_1-\phi_2)}.
\end{eqnarray}
In our case of $D\to \pi^+\pi^-\; (\pi^0\pi^0)$, the two interfering
amplitudes are the modified ``tree" part $\tilde T^{tree}(D\to\pi\pi)$
(\ref{tilde1}) and the pure ``inelastic FSI" part $T^{inelastic}$ given
in (\ref{ine}), which have different weak phases $\delta_i$ 
and strong phases $\phi_i$.
For $D^0\to\pi^0\pi^0$ the expressions are similar.

Let us work within the framework of CKM matrix \cite{Data}
\begin{eqnarray}
V_{ud}V_{cd}^* &=& (c_{12}c_{13})(-s_{12}c_{23}-c_{12}s_{23}s_{13}
e^{i\delta_{13}})^* \\
V_{us}V_{cs}^* &=& (s_{12}c_{13})(c_{12}c_{23}-s_{12}s_{23}s_{13}
e^{i\delta_{13}})^*.
\end{eqnarray}
Thus
\begin{eqnarray}
\delta^{D\to \pi} &=& arctan{-c_{12}s_{23}s_{13}\sin\delta_{13}
\over s_{12}c_{23}+c_{12}s_{23}s_{13}\cos\delta_{13}};\\
\delta^{D\to K} &=& arctan{s_{12}s_{23}s_{13}\sin\delta_{13}
\over c_{12}c_{23}-s_{12}s_{23}s_{13}\cos\delta_{13}}.
\end{eqnarray}

In $\tilde T^{tree}(D^0\to\pi^+\pi^-)\equiv
T^{tree}(D^0\to\pi^+\pi^-)+T^{FSI}(D^0\to\pi\pi\to
\pi^+\pi^-)$, the weak phase is that of the tree amplitude as
$\delta^{D\to\pi}$, while that of $T^{FSI}(D^0\to
K^+K^-\to \pi^+\pi^-)$ is $\delta^{D\to K}$,
therefore they are different.
One can also notice that $|\delta^{D\to K}|\ll |\delta^{D\to
\pi}|$ in this convention, even though the final result is independent of
the convention adopted in the calculations.

Later we will evaluate
\begin{eqnarray}
R_1 &\equiv& {\Gamma(D^0\to\pi^+\pi^-)-\Gamma(\bar D^0\to
\pi^+\pi^-)\over \Gamma(D^0\to\pi^+\pi^-)+\Gamma(\bar D^0\to
\pi^+\pi^-)}; \\
R_2 &\equiv & {\Gamma(D^0\to\pi^0\pi^0)-\Gamma(\bar D^0\to
\pi^0\pi^0)\over \Gamma(D^0\to\pi^0\pi^0)+\Gamma(\bar D^0\to
\pi^0\pi^0)}.
\end{eqnarray}
In next section, we will present our numerical results of $R_1$ and
$R_2$. \\

\noindent{\bf III. The numerical  results}
\vspace{0.3cm}

(i) All the CKM entries involved in our calculations have been measured,
even though there are some uncertainties \cite{Data}. Thus with
the optimistic $\sin\delta_{13}=1$, we have
$$\delta^{D\to\pi}\simeq 3.2\times 10^{-4}\sim 1.1\times 10^{-3},\;\;\;\;
\delta^{D\to K}\simeq 5.3\times 10^{-5}.$$
Later we will use the most favorable values for the CP violation
calculations.

(ii) We obtain in terms of the dipole approximation,
$$f_+^{D\pi}(q^2=m_{\pi}^2)\approx 0.8;\;\;\; f_-^{D\pi}(q^2=m_{\pi}^2)
\approx -0.8; $$
$$f_+^{D}(q^2=m_{K}^2)\approx 0.7;\;\;\; f_-^{DK}(q^2=m_{K}^2)
\approx -0.65. $$

(iii) For the elastic and inelastic scattering $\pi\pi\rightleftharpoons
\pi\pi$, $K\bar K\rightleftharpoons \pi\pi$,
the transition probability amplitude $T$ and the phase shift $\theta$ are
experimentally measured \cite{BSZ} as
$$T_0(\pi\pi\rightleftharpoons\pi\pi)\approx -0.48,\;\;{\rm and}\;\; \theta_0
\approx 308^{\circ},$$
$$T_2(\pi\pi\rightleftharpoons\pi\pi)\approx -0.45,\;\;{\rm and}\;\; \theta_2
\approx -50^{\circ};$$
$$T^{inelastic}=T_K(K\bar K\rightleftharpoons\pi\pi)\approx 0.1,
\;\;{\rm and}\;\;\theta_K(K\bar K\to\pi\pi)\approx 310^{\circ},$$
for the energy range of $M_D$. So by the notation of eq.(\ref{dircp})
$\sin(\delta_1-\delta_2)\sim -1.1\times 10^{-3}, $
while $\phi_2\approx 310^{\circ}$, but $\phi_1$ of eq.(\ref{dircp})
must be evaluated by eq.(\ref{tilde1}) and (\ref{tilde2}),
in our case.

(iv) The CP asymmetries.

With the information given above, we obtain
\begin{eqnarray}
R_1 &=& -1.1\times 10^{-4};\\
R_2 &=& 2.2\times 10^{-4}.
\end{eqnarray}
In these calculations we almost do not have any free parameters, except the
CKM phase $\delta_{13}$. We have taken $sin\delta_{13}=1$ and neglected
the contribution from other intermediate states. They may modify the
results, but are unlikely to change the 
whole scenario and order of magnitude of the CP asymmetry.
It is noted that $R_1$ and $R_2$ have opposite signs, as a matter
of fact, due to the uncertainty of $\delta_{13}$ in the CKM matrix, the
absolute sign of $R_i$ is not important, but only the relative sign
is meaningful.\\

\noindent {\bf IV. Conclusion and discussion}
\vspace{0.3cm}

In this work, we discuss the CP violation effects caused by the inelastic
FSI rescattering in $D\to\pi^+\pi^-$, $\pi^0\pi^0$
modes and obtain the CP asymmetry ratios of order $10^{-4}$.

(i) The observed indirect CP violation in the K-system which is characterized
by $\epsilon$ is of order $10^{-3}$. Even though the direct CP violation
$\epsilon'/\epsilon$ has not been reliably measured yet, present data
incline to confirm it to be $(1.5\pm 0.8)\times 10^{-3}$.
Anyhow, one has all reasons to
believe it is non-zero and of order $10^{-4}\sim 3\times 10^{-3}$
\cite{Data}, namely the superweak mechanism is almost ruled out by experiments.
Our estimation on $D\to\pi\pi,\; K \bar K$ shows that the direct CP
asymmetries here are about 2 orders of magnitude larger than $\epsilon'$.

(ii) In the expression, one can see that the direct CP violation is caused
by the interference between the tree amplitude modified by the elastic
rescattering (denoted as $\tilde T^{tree}$ in this paper) 
and that induced by the
inelastic FSI rescattering while the two parts have different weak and strong
phases. The interference is proportional to a product of the two amplitudes
$|\tilde T^{tree}||T^{FSI}|$ and the differences of weak and strong angles
$|\sin(\delta^{D\pi}-\delta^{DK})\cdot\sin(\phi_1-\phi_2)|$.
There are two factors
which suppress the CP asymmetry values.

The first one is that in the framework of CKM matrix, the weak phase
is about order $10^{-3}$ which is independent of convention.

The second suppression factor comes from 
the measured strong phase difference $|\sin(\phi_1-\phi_2)|=0.3$ and 
amplitude $T_K(K\bar K\to\pi\pi)\sim 0.1$. 

As discussed in ref.\cite{XQ,Zou}, the final state interaction effects can be
described in a hadronic triangle diagram and the absorptive part of the
loop gives rise to a strong phase. In our previous work, we only estimated the
absorptive part and stressed that the inelastic FSI is important in many
processes, so that cannot be ignored. The real part is hard to be properly
evaluated because of the ultraviolet divergence and the obtained results
would depend on the renormalization scheme.
In this work, as suggested in literatures \cite{Chau1}, we only deal with the
FSI, namely the intermediate hadrons are real particles, i.e. on their mass
shell, the dispersive part of the loop is small compared to the tree
amplitude, in fact, its contribution is effectively absorbed into the
phenomenological parameters $a_1, \; a_2$ which may slightly deviate from
the values derived in terms of the renormalization group equation, in our
case (Note, it is not always true).
Thus, we use only
the data directly obtained from corresponding experiments, so can avoid
any ambiguity caused by theoretical uncertainties.

(iii) Since the proposed channels to be observed are Cabibbo-suppressed,
the decay rates could be smaller than the Cabibbo favored channels by
$\sin^2\theta_C$ roughly.

(iv)
The branching ratio of $D\to K\bar K;\pi\pi$
is about $2\times 10^{-3}$.
The production cross section $\sigma$ of $D^0\bar D^0$
is measured at BEPC \cite{Yan},
$$\sigma(D^0\bar D^0)=11.63\pm 1.1\;nb,\;\;\; {\rm at\; the\; BEPC\; energy}.$$
Taking the most optimistic values evaluated in the framework of
C-K-M theory, the number of events for observing
the Cabibbo-suppressed decay channels $D\to K\bar K;\pi\pi$ would be
\begin{equation}
N=L\times 2\times 10^{-3}\times\sigma\times\tau\times f,
\end{equation}
where L is the luminosity, $\tau$ is the measuring time period and f is the
observation efficiency.

For the proposed charm-tau factory, L can reach $10^{34}\; cm^{-2}
sec^{-1}$, so
$$N\approx 7.3\times 10^{6}\times f\times n,$$
where n is the number of necessary years.
Since the
CP asymmetry is ranged about $10^{-4}$, so to
the reasonable statistical level for observation of CP violation,
N at least must be $10^{7}\sim 10^{8}$,
it would need a charm-tau factory  with luminosity of $10^{34}cm^{-2}sec^{-1}$
to run for 15 years. Even though this number is not much encouraging,
but as suggested by Browder and Pakvasa \cite{Brow}, if there is new physics
which can provide with a larger weak phase, the observation becomes very
possible. Even with this small CP asymmetry, there is still possibility
to make the measurement. 

Therefore, for measuring the direct CP violation which is one of the main
interests in the field of high energy physics, a high luminosity $\tau$-charm
factory would be extremely helpful.\\

\noindent{\bf Acknowledgments}

One of us (Li) would like to thank Dr. X.-G. He for helpful discussions.\\

\end{document}